\documentclass[twocolumn,aps,prl,superscriptaddress,showpacs]{revtex4}
\usepackage{graphicx}
\usepackage{amsmath}
\usepackage{epsfig}
\usepackage[utf8]{inputenc}

\begin{document}
 \title{Demonstration of Coherent State Discrimination Using a Displacement Controlled Photon Number Resolving Detector}

\author{Christoffer Wittmann}
\affiliation{Max Planck Institute for the Science of Light, G\"{u}nther-Scharowsky-Str. 1, Bau 24, 91058, Erlangen, Germany}
\affiliation{Institut f\"{u}r Optik, Information und Photonik, University of Erlangen-Nuremberg, Staudtstra\ss e 7/B2, 91058, Erlangen, Germany}

\author{Ulrik L. Andersen}
\affiliation{Department of Physics, Technical University of Denmark, Building 309, 2800 Kgs. Lyngby, Denmark}
\affiliation{Max Planck Institute for the Science of Light, G\"{u}nther-Scharowsky-Str. 1, Bau 24, 91058, Erlangen, Germany}
\affiliation{Institut f\"{u}r Optik, Information und Photonik, University of Erlangen-Nuremberg, Staudtstra\ss e 7/B2, 91058, Erlangen, Germany}

\author{Masahiro Takeoka}
\affiliation{National Institute of Information and Communications Technology (NICT),
4-2-1 Nukui-kitamachi, Koganei, Tokyo 184-8795, Japan}

\author{Denis Sych}
\affiliation{Max Planck Institute for the Science of Light, G\"{u}nther-Scharowsky-Str. 1, Bau 24, 91058, Erlangen, Germany}

\author{Gerd Leuchs}
\affiliation{Max Planck Institute for the Science of Light, G\"{u}nther-Scharowsky-Str. 1, Bau 24, 91058, Erlangen, Germany}
\affiliation{Institut f\"{u}r Optik, Information und Photonik, University of Erlangen-Nuremberg, Staudtstra\ss e 7/B2, 91058, Erlangen, Germany}

\date{\today}

\begin{abstract}
We experimentally demonstrate a new measurement scheme for the discrimination of two coherent states. The measurement scheme is based on a displacement operation followed by a photon number resolving detector, and we show that it outperforms the standard homodyne detector which we, in addition, proof to be optimal within all Gaussian operations including conditional dynamics. We also show that the non-Gaussian detector is superior to the homodyne detector in a continuous variable quantum key distribution scheme.
\end{abstract}

\pacs{03.67.Hk, 03.65.Ta, 42.50.Lc}

\maketitle
According to the basic postulates of quantum mechanics, perfect discrimination of non-orthogonal quantum states is impossible. Suppose, for example, one is randomly given one of two a priori
known coherent states, then there is no physical apparatus that with certainty
can identify which state was at hand due to the intrinsic non-orthogonality of coherent states. This inability to perfectly discriminate coherent states is the engine for unconditionally secure
communication via continuous variable quantum key distribution~\cite{grosshans_quantum_2003}. On the other hand, in order to increase the secure key rate, optimized discrimination strategies must be implemented. Optimized measurements for coherent state discrimination are also of great use in other quantum devices such as quantum computers~\cite{Ralph2003} and quantum repeaters~\cite{van_loock_hybrid_2006}.

The impossibility of perfectly discriminating quantum state has therefore lead to the fundamental problem of finding measurement strategies for which the discrimination task is optimised with respect to different figure of merits. The two most well-known discrimination strategies are deterministic minimum error state discrimination (MESD) and  probabilistic unambigious state discrimination (USD)~\cite{helstrom_quantum_1976,ivanovic_to_1987,dieks_overlap_1988, peres_to_1988, jaeger_optimal_1995}. In an optimised MESD measurement all measurement outcomes are kept and the error rate is minimized whereas in an optimised USD measurement only conclusive measurement outcomes are kept while the rate of inconclusive results is minimized. Experimental realizations of such measurement strategies have been pursued~\cite{Riis1997,huttner_unambiguous_1996, barnett_quantum_2009, cook_optical_2007, bartkov_programmable_2008,wittmann_demonstration_2008}.

A combination of the two discrimination schemes - the intermediate discrimination (ID) scheme - where one allows for both erroneous and inconclusive results has also been treated theoretically. More precisely, the minimal probability of errors for a fixed probability of inconclusive results has been derived for pure and mixed states in refs.~\cite{chefles_strategies_1998} and~\cite{fiurek_optimal_2003}, respectively. A well-known (however non-optimal) ID scheme is the postselection based homodyne detector where the quadrature measurement outcomes are favourably postselected. Such a measurement scheme has been used to discriminate binary coherent states in quantum key distribution~\cite{silberhorn_continuous_2002, lorenz_continuous-variable_2004, lance_no-switching_2005} but also to discriminate noisy non-classical states for distillation~\cite{heersink_distillation_2006,dong_experimental_2008,franzen_experimental_2006,hage_preparation_2008} and to engineer quantum state\cite{marek_resources_2009,lance_quantum-state_2006}. 

In this Letter, we first show theoretically that the postselection based homodyne detector is the optimal intermediate discrimination strategy for binary coherent states over all Gaussian measurement approaches (including Gaussian transformations, homodyne detectors and conditional dynamics). Furthermore, we experimentally implement a Non-Gaussian measurement strategy (based on a displacement controlled photon number resolving (PNR) detector) that outperforms the optimal Gaussian strategy. Finally, we show theoretically that by using the non-Gaussian detector in replacement of the postselection based homodyne detector in a continuous variable quantum key distribution protocol, a substantial increase of the secure key rate is expected.  

Consider a binary alphabet of two pure and phase
shifted coherent states $\{|{-}\alpha\rangle,|\alpha\rangle\}$ occuring with the a
priori probabilities $p_1$ and $p_2$. The task of the receiver is
to certify whether the state was prepared in $|{-}\alpha\rangle$ or $|\alpha\rangle$ using a measurement described by the three-component positive operator-valued measure (POVM) $\hat\Pi_i, i=1,2,?$ where $\hat\Pi_i {>}0$ and $\hat\Pi_1{+}\hat\Pi_2{+}\hat\Pi_?{=}\hat I$. An inconclusive result will occur with the probability
\begin{equation} 
p_\mathrm{inc}=p_1\langle {-}\alpha|\hat\Pi_?| {-}\alpha\rangle+p_2\langle \alpha|\hat\Pi_?|\alpha\rangle,
\end{equation}
where $\langle {-}\alpha|\hat\Pi_?|{-}\alpha\rangle$ ($\langle \alpha|\hat\Pi_?|\alpha\rangle$) represents the  probability of inconclusive results when $|{-}\alpha\rangle$ $(|\alpha\rangle)$ was prepared.
Furthermore, the average error probability is given by
\begin{equation}
p_\mathrm{E}=(p_1\langle{-}\alpha|\hat\Pi_2|{-}\alpha\rangle+p_2\langle \alpha|\hat\Pi_1|\alpha\rangle)/(1{-}p_\mathrm{inc}), 
\end{equation}
where $\langle{-}\alpha|\hat\Pi_2|{-}\alpha\rangle$ ($\langle\alpha|\hat\Pi_1|\alpha\rangle$) represents the error probability of mistakenly guessing $|{-}\alpha\rangle$ $(|\alpha\rangle)$. An optimised intermediate detector has a minimal error probability, $p_\mathrm{E}$, for a given probability of inconclusive results, $p_\mathrm{inc}$.

\begin{figure}
\begin{tabular}{l}
\centerline{\includegraphics[width=7cm]{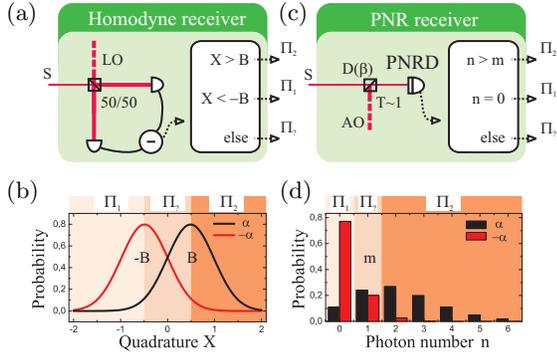}} \\ [-4.8cm]
\hspace{0.5cm}(a)\hspace{3.2cm}(c) \\[2cm]
\hspace{0.5cm}(b)\hspace{3.2cm}(d) \\[1.9cm]
\end{tabular}
\caption{\label{schemes} (a) Schematics of the homodyne receiver. The signal (S) is interfered with a local oscillator (LO). The photocurrents are subtracted resulting in a quadrature measurement. (b) Marginal distribution of the two signal states with intervals where the answers $\lbrace -, ?, + \rbrace$ are guessed. (c) Schematics of the photon number resolving (PNR) receiver. The signal (S) is interfered with an auxiliary oscillator (AO). Finally, the signal is measured by a photon number resolving detector (PNRD) (d) Photon number distribution of two signal states. In the examples, we assume a signal with $|\alpha|^2=0.24$ and a displacement of $\beta=1$.
\vspace{-0.5cm}}
\end{figure}

\begin{figure}
\begin{tabular}{l}
\centerline{\includegraphics[width=7cm]{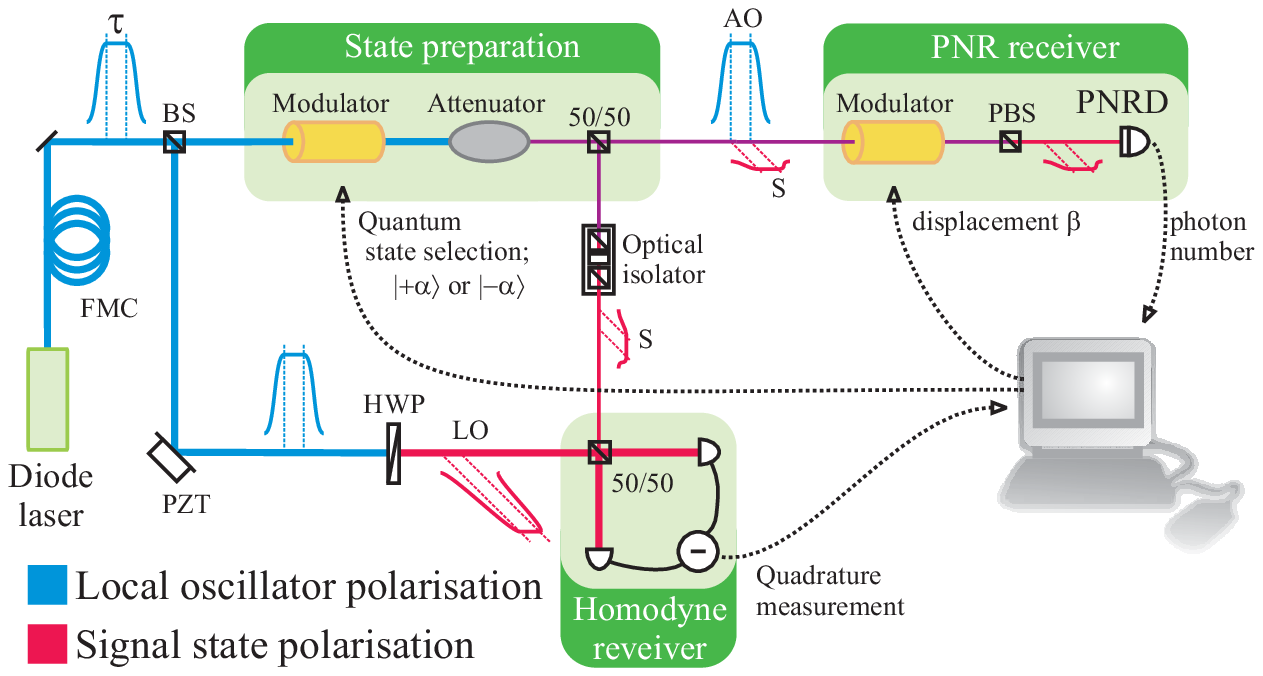}}\\[0.1cm]
\centerline{\includegraphics[width=7cm]{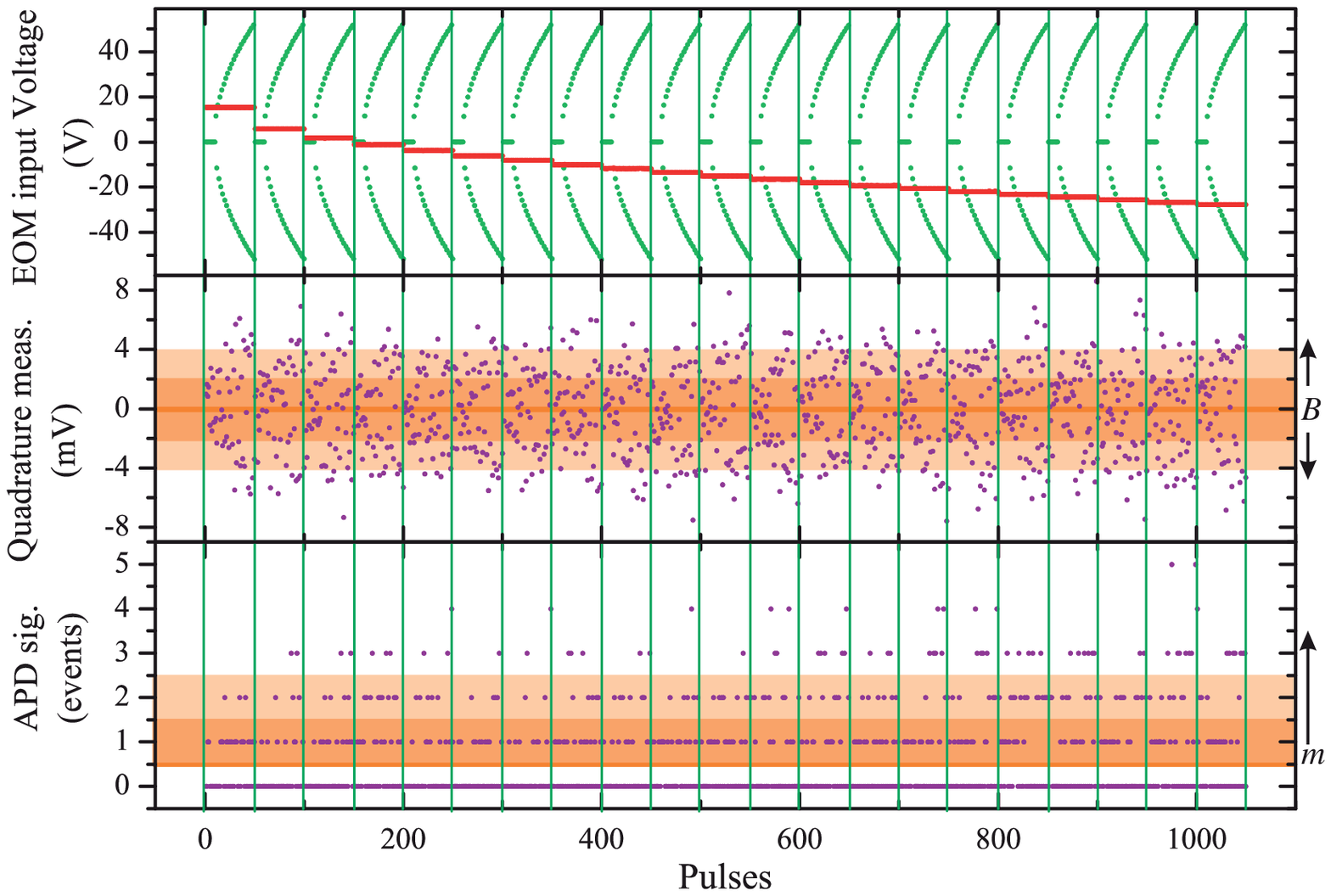}}\\[-8.8cm]
(a)\\[3.5cm]
(b)\\[4.2cm]
\end{tabular}
\caption{\label{SetupRaw} (a) Scheme of the experimental implementation of the receivers in Figs.~\ref{schemes}(a,c), where the abbreviated components are a fibre mode cleaner(FMC), beam splitters (BS, 50/50), a polarizing beam splitter (PBS), a piezo mounted mirror (PZT), a half wave plate (HWP) and a photon number resolving detector (PNRD). (b top) Modulation pattern of the electro optical modulators in the state preparation (green) and the displacement of the PNR receiver (red), (b middle) simultaneously recorded quadrature measurements and (b bottom) detection events of the APD. Shaded areas show inconclusive results for increasing postselection parameters $B$ and $m$.
\vspace{-0.5cm}}
 
\end{figure}

An experimentally simple candidate for an intermediate measurement is the homodyne detector measuring the excitation quadrature followed by postselection of the outcomes as illustrated in Fig.~\ref{schemes}(a)~\cite{silberhorn_continuous_2002, lorenz_continuous-variable_2004, lance_no-switching_2005}. The  distribution of the measurement outcomes is shown in  Fig.~\ref{schemes}(b) and is divided into three regions associated with the three POVM elements. If the measurement outcome is larger (smaller) than a certain threshold value, $B (-B)$, then we identify $|\alpha\rangle (|-\alpha\rangle)$ (with errors), otherwise the outcomes are inconclusive. The trade-off between the error probability and the probability of inconclusive results can be easily tuned by the threshold value, $B$. 

In the following we prove that this measurement scheme is the optimal 
strategy for realizing the intermediate measurement within all possible 
Gaussian operations and conditional dynamics (classical feedback or 
feedforward).
First we note that if the input alphabet as well as all operations 
are Gaussian, conditional dynamics is 
useless~\cite{giedke_characterization_2002}. 
In our case, however, the input alphabet consisting of an ensemble of 
two coherent states is non-Gaussian, and thus we cannot discard conditional 
dynamics as an option to improve the discrimination task. 
The POVM consisting of noise-free Gaussian operations 
without conditional dynamics is described by a set of operators 
$\{ \frac{1}{\pi}|\psi_\zeta (u,v)\rangle\langle\psi_\zeta (u,v)| \}$ 
where $|\psi_\zeta(u,v)\rangle = \hat{D}(u,v) \hat{S}(\zeta) |0\rangle$ 
is a displaced squeezed state, 
$\zeta=re^{i\varphi}$ is a complex squeezing parameter, 
and $(u,v)$ are quadratures representing a measurement outcome 
\cite{takeoka_discrimination_2008}. 
The probability 
distributions of detecting $|{\pm}\alpha\rangle$ with this POVM are 
$P(u,v|{\pm}\alpha){=}\frac{1}{\pi}|\langle\pm\alpha|\psi_\zeta(u,v)\rangle|^2$
showing Gaussian statistics. 
Let us denote the likelihood ratio of two signals as 
$\Lambda_1{=}\frac{p_1 P(u,v|{-}\alpha)}{p_2 P(u,v|\alpha)}$ 
and $\Lambda_2{=}\Lambda_1^{-1}$. 
According to the Bayesian strategy \cite{helstrom_quantum_1976}, 
an optimal signal decision for the fixed measurement is to guess 
$|{-}\alpha\rangle$ for $\Lambda_1 {\ge} \Lambda_B$, 
$|\alpha\rangle$ for $\Lambda_2 {\ge} \Lambda_B$, 
and the inconclusive result otherwise, 
where $\Lambda_B$ is the threshold. 
The error probabilities and the probabilities of having inconclusive result 
for each signal are then given by 
$p_e^{(\pm)} {=} \frac{1}{2} {\rm erfc}\left[ \sqrt{2a}\alpha 
{+} \frac{\ln\Lambda_B \mp \ln(p_1/p_2)}{4\sqrt{2}\alpha} \right]$, 
where 
$p_i^{(\pm)} {=} p_s^{(\pm)} {-} p_e^{(\pm)}$, 
$p_s^{(\pm)} = \frac{1}{2} {\rm erfc}\left[ \sqrt{2a}\alpha 
{-} \frac{\ln\Lambda_B \pm \ln(p_1/p_2)}{4\sqrt{2}\alpha} \right]$, 
$a=\frac{1+\cosh 2r+\sinh 2r \cos\varphi}{2( \cosh 2r+1 )}$, 
and $\alpha$ is assumed to be real and positive for simplicity. 
Here we can find that 
the average error probability $p_E {=} (p_1 p_e^{(-)} {+} p_2 p_e^{(+)})
/(1{-}p_{\rm inc})$ and the inconclusive probability 
$p_{\rm inc} {=} p_1 p_i^{(-)} {+} p_2 p_i^{(+)}$ are simultaneously 
minimized with $\varphi=0$ and $r\to\infty$. 
It corresponds to an X-quadrature measurement, 
implying that the optimal measurement with only Gaussian operations 
is the homodyne detector with phase $\varphi {=} 0$.

Furthermore, any conditional operation 
is proven to be useless by considering two Gaussian operations.
The first Gaussian operation on the input state 
includes a partial measurement of the signal and generally outputs 
a measurement outcome and a conditional output state. 
It was shown that such a conditional state can always be transformed into 
another mixture of coherent states 
$\hat{\rho}_{out} {=} p'_1(d_{\mathcal{M}}) 
|$$-$$\alpha'\rangle\langle -\alpha'| 
{+} p'_2(d_{\mathcal{M}}) |\alpha'\rangle\langle\alpha'|$, 
with real and positive $\alpha'$,  
by an additional Gaussian operation which is deterministic and 
independent of the partial measurement outcome denoted by $d_{\mathcal{M}}$ 
\cite{takeoka_discrimination_2008}. 
Since only the posteriori probabilities depend on $d_{\mathcal{M}}$, 
the optimal second operation 
is independent of $d_\mathcal{M}$ and given by 
a fixed homodyne measurement ($\varphi{=}0$) as already shown. 
We therefore conclude that any conditional dynamics is not useful 
in the two-step measurement scenario. 
An extension of the above conclusion to the multi-step measurement scenario 
is straightforward, which proves the optimality of the homodyne detector 
within all possible Gaussian operations and conditional dynamics~\cite{wittmann_preparation_2009}.

%

Although the homodyne detector is optimal within all Gaussian strategies, there exist non-Gaussian strategies that out-perform the homodyne detector. In the following we discuss a new, non-Gaussian discrimination detector that beats the performance of the homodyne detector. It is based on a displacement operation, $D(\beta)$, followed by a detection of the photon number (see Fig.~\ref{schemes}(c)) with which conclusions are made~\cite{wittmann_discrimination_2009}. The photon number distributions of the two possible coherent states after displacement are shown in Fig.~\ref{schemes}(d). For zero photon outcomes, we identify $|-\alpha\rangle$ (since the zero-photon contribution from $|-\alpha\rangle$ is much larger than from $|\alpha\rangle$) and associate the POVM, $\hat\Pi_1{=}|0\rangle\langle 0|$. If the photon number outcome, $n$, is larger than a certain threshold, $m$, we identify $|\alpha\rangle$ with the POVM, $\hat\Pi_2{=}\hat I {-}\hat\Pi_1{-}\hat\Pi_?$, otherwise we obtain inconclusive results described by the POVM, $\hat\Pi_?{=}\sum_{n=1}^{m}{|n\rangle\langle n|}$. To minimize the error rate the displacement must be optimized, and a detailed discussion on this optimization procedure can be found in Ref.~\cite{wittmann_discrimination_2009}.

\begin{figure}
\begin{tabular}{l}
\centerline{\includegraphics[width=7.6cm]{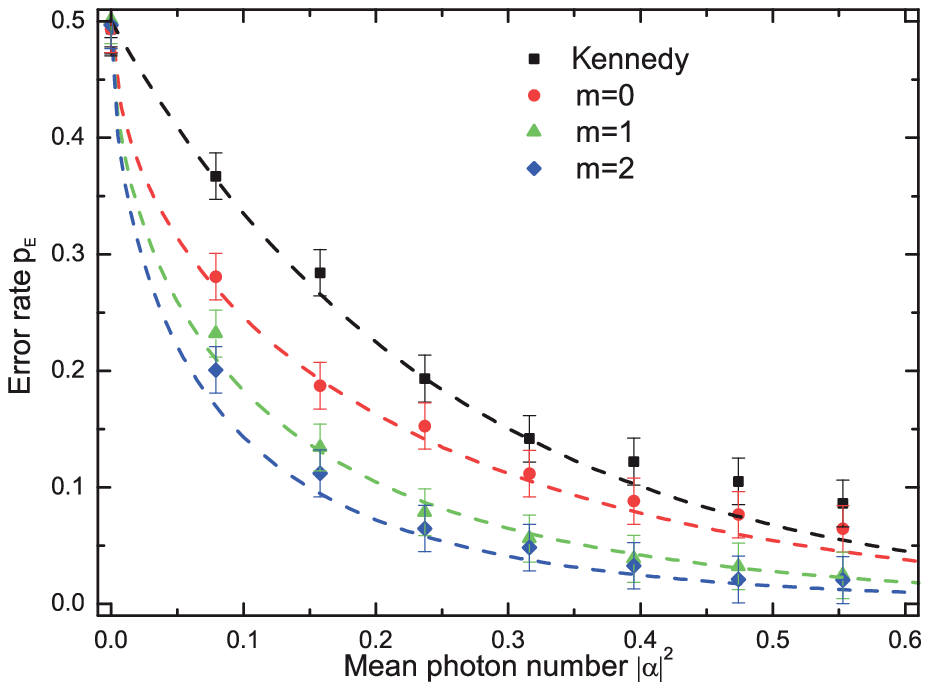}}\\[0.2cm]
\centerline{\hspace{0.2cm}\includegraphics[width=7.4cm]{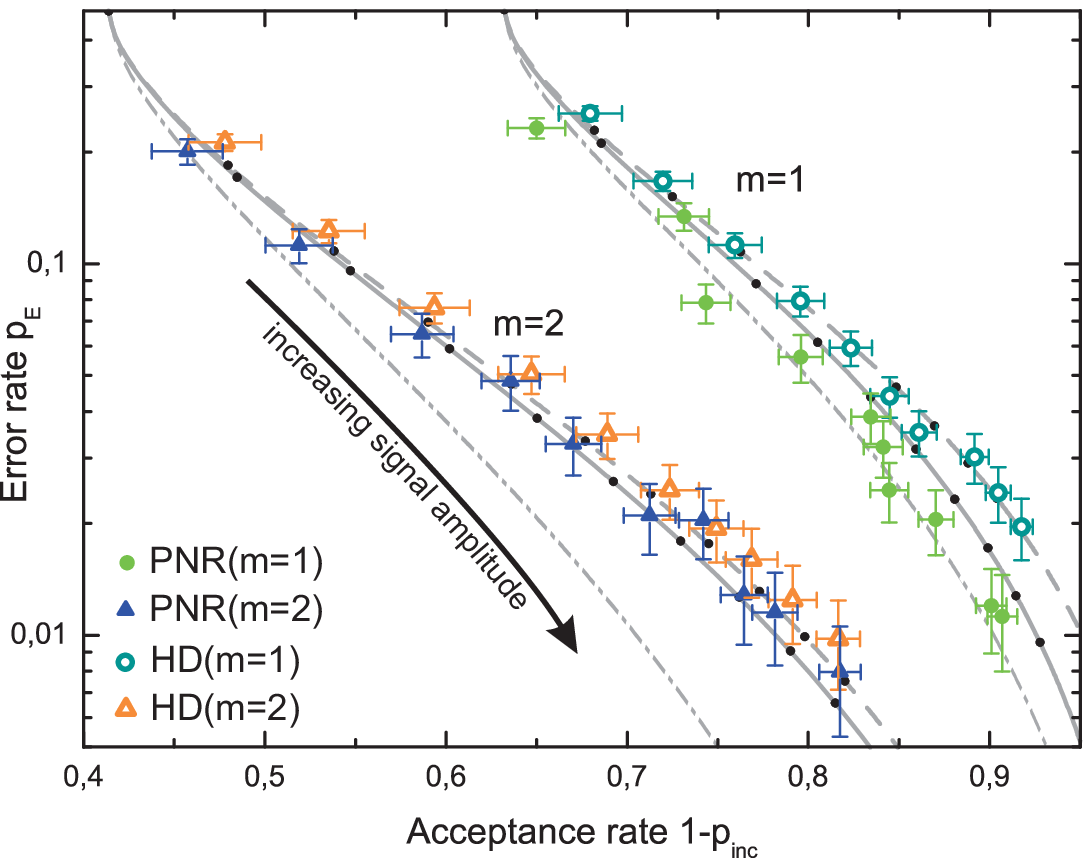}}\\
\centerline{\hspace{0.2cm}\includegraphics[width=8cm]{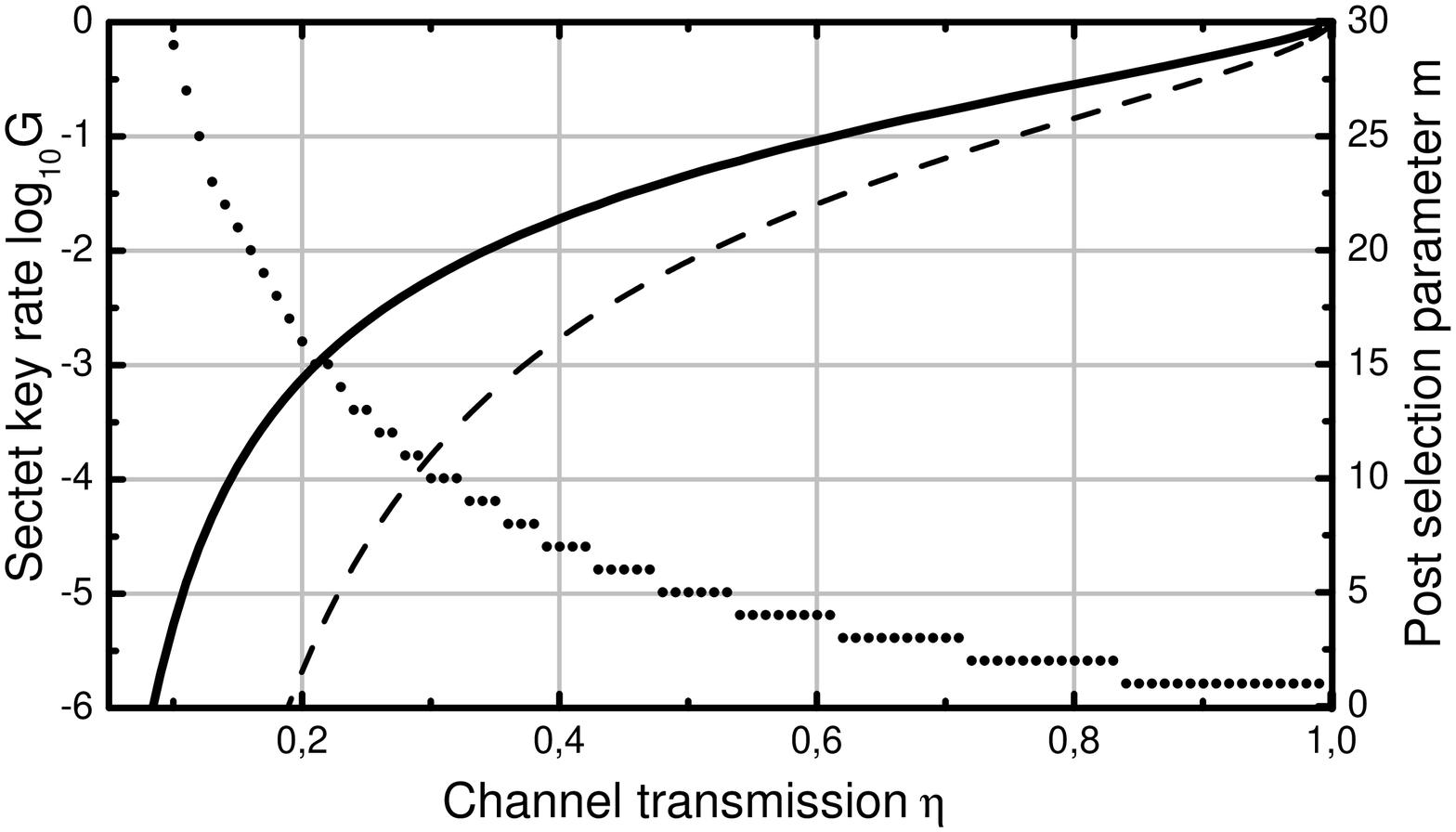}}\\[-16.5cm]
(a)\\[5.5cm]
(b)\\[5.6cm]
(c)\\[4.0cm]
\end{tabular}

\caption{\label{comprates} (a) Error rates for the Kennedy receiver ($\beta=\alpha$) and the PNR receiver with varying $m$ and optimized $\beta_\mathrm{opt}$. Error bars reflect the standard deviations of repeated measurements, which are larger than the statistical errors. Experimental data is compared to ideal receivers (dashed lines). The experimental data has been corrected for the detector inefficiency in all figures. (b) Experimental error rates versus acceptance rates with increasing signal amplitudes for the PNR receiver and the homodyne receiver. For this comparison the success rate of both schemes is fixed to the one, that is theoretically reached by the PNR receiver. The theoretical predictions for the homodyne receiver (grey dashed line), the PNR receiver (solid line) and the optimal intermediate measurement (dotdashed lines) are shown with statistical error bars. (c left scale) Key rate $G$ in logarithmic scale as a function of the channel transmittance $\eta$ using the PNR receiver (solid curve) and the homodyne receiver \cite{heid_efficiency_2006,sych_coherent_2009} (dashed curve).
(c right scale) Optimized threshold $m$ (dotted line). Photon number resolution for high photon number, e.g. $m{=}10$, was demonstrated in~\cite{afek_quantum_2009}.}
 
\end{figure}

In Fig.~\ref{comprates}(b)(grey lines) we compare our new detector with the homodyne receiver by choosing the postselection parameter $B$ such that the rates of inconclusive results are equal for both strategies, i.e. $p_\mathrm{inc,HD}{=}p_\mathrm{inc,PNR}$. We find that the displacement controlled PNR detector (solid lines) surpasses the performance of the homodyne detector (dashed lines) for all signal amplitudes.

We continue with a description of the experimental realization of the two detector schemes. As shown in Fig.~\ref{SetupRaw}(a) the setup consists of a preparation stage and two different receiver stages. The signal states are generated in a polarisation mode orthogonal to an auxiliary mode: The field amplitude of the auxiliary mode is coherently transfered into the signal polarisation by means of an electro-optical modulator (EOM). We carefully characterize the prepared signal and both detectors and verify that the amount of excess noise stemming from the EOM is miniscule~\cite{wittmann_demonstration_2008}. Using a 50/50 beam splitter, two identical signal states are directed to the two detection schemes. The homodyne receiver records a quadrature value for each signal pulse. Its quantum efficiency amounts to $\eta_{\mathrm{hom}}=85.8\%$, and the electronic noise level is more than $23\:\mathrm{dB}$ below the shot noise level.
The PNR receiver is composed of a displacement operation (driven by an EOM and the auxiliary mode) and a fiber coupled avalanche photo diode (APD) operating in an actively gated mode, such that the dead time of the device ($50\,$ns) is much shorter than the measurement time ($800\,$ns) (which defines the duration of the state). The APD therefore works as a primitive photon number resolving detector~\cite{banaszek_direct_1999}. The interference of the signal and auxiliary oscillator is performed with an extinction ratio of about 1/700 and the total detection efficiency of the displacement operation and the detection is estimated to be $\eta_{\mathrm{on/off}}=55\%$. 
A PC controls the preparation of the states as well as the displacement in the PNR receiver by modulating the two EOMs. Simultaneously it acquires the homodyne and the APD detection outcomes during the pulse sequence. (The quadrature values are derived by averaging 16 samples of homodyne data (bandwidth 10\,MHz), which is digitized with 20MS/s. The detected temporal modes at both receivers are therefore equal to a reasonable extent.) An example of such a sequence is shown in Fig.~\ref{SetupRaw}(b). The outcomes of the receivers are then divided into correct, false and inconclusive results.

In our experiment the PNR receiver is demonstrated for $m=0$, $1$ and $2$. We find that for any $m$ the displacement can be optimized such that the experimentally measured error rates reach a minimum. The optimal displacement is higher for higher $m$ and the minimum error rate after this optimization of the displacement is lowered for increasing $m$. The error rates for varying amplitudes are plotted in Fig.~\ref{comprates}(a). We find a maximal reduction of the error rate by a factor of $3.5$ going from $m=0$ (deterministic scheme) to $m=2$ (probabilistic scheme) at the signal $|\alpha|^2=0.47$.
The corresponding penalty on the acceptance rates and the comparison with the theoretical predictions are shown in Fig.~\ref{comprates}(b). In this figure the performance of the homodyne detector is compared with the performance of the displacement controlled PNR detector and it can be clearly seen (especailly for $m=1$) that the latter non-Gaussian detector outperforms the former Gaussian detector at several data points. For details see~\cite{wittmann_preparation_2009}.

In the last part of the letter, we investigate the performance of a continuous variable quantum key distribution scheme in which the standard homodyne detector is replaced by an ideal displacement controlled PNR receiver at the receiving station. We consider a binary coherent state alphabet, a lossy channel (with no excess noise) and an error correction scheme based on direct reconciliation ~\cite{heid_efficiency_2006}. For this scheme the secret key rate is $G=(1-p_\mathrm{inc})(I_{B}-I_{E})$ with Bob's information $I_{B}=1-H(p_\mathrm{err})$, and it depends on the channel transmittance $\eta$, the signal amplitude $\alpha$, the displacement value $\beta$, and the threshold value $m$.
We calculate the key rate $G$ as a function of the channel transmittance $\eta$ while optimizing the other parameters (typically $\alpha{\in}[0.5{,}1.5]$). The result is shown in Fig.~\ref{comprates}(c) (solid curve). For comparison we also insert the key rate for the standard homodyne detection based protocol (dashed curve)~\cite{sych_coherent_2009}. We find that the new scheme is far superior to the homodyne scheme by several orders of magnitude, especially in the realistic case of high channel attenuation. 

In this Letter, we have experimentally demonstrated a receiver for binary-encoded optical coherent states based on an optimized displacement, a photon number resolving measurement and postselection of the measured photon number outcomes. We compared this receiver scheme to the homodyne receiver, which we proved to be the optimal Gaussian receiver, and found that the performance of the PNR receiver beats that of the homodyne receiver. The quantum efficiency for both receivers is approaching unity due to rapid development in this field. Consequently, we showed theoretically that by using an ideal version of the new receiver in replacement of the standard homodyne receiver in a continuous variable QKD protocol, superior performance in terms of increased secure key rate is expected. QKD is just one application among many others for which the PNR receiver demonstrates superior performance in comparison with a homodyne based scheme, and thus we believe that our new detector will play a significant role in future quantum information technologies.

The work has been supported by the EU project COMPAS and Lundbeckfonden(R13-A1274).

%

\begin{thebibliography}{31}
\expandafter\ifx\csname natexlab\endcsname\relax\def\natexlab#1{#1}\fi
\expandafter\ifx\csname bibnamefont\endcsname\relax
  \def\bibnamefont#1{#1}\fi
\expandafter\ifx\csname bibfnamefont\endcsname\relax
  \def\bibfnamefont#1{#1}\fi
\expandafter\ifx\csname citenamefont\endcsname\relax
  \def\citenamefont#1{#1}\fi
\expandafter\ifx\csname url\endcsname\relax
  \def\url#1{\texttt{#1}}\fi
\expandafter\ifx\csname urlprefix\endcsname\relax\def\urlprefix{URL }\fi
\providecommand{\bibinfo}[2]{#2}
\providecommand{\eprint}[2][]{\url{#2}}

\bibitem[{\citenamefont{Grosshans et~al.}(2003)\citenamefont{Grosshans, Assche,
  Wenger, Brouri, Cerf, and Grangier}}]{grosshans_quantum_2003}
\bibinfo{author}{\bibfnamefont{F.}~\bibnamefont{Grosshans}},
  \bibnamefont{et~al.}, \bibinfo{journal}{Nature}
  \textbf{\bibinfo{volume}{421}}, \bibinfo{pages}{238} (\bibinfo{year}{2003}).

\bibitem[{\citenamefont{Ralph et~al.}(2003)\citenamefont{Ralph, Gilchrist,
  Milburn, and Glancy}}]{Ralph2003}
\bibinfo{author}{\bibfnamefont{T.~C.} \bibnamefont{Ralph}},
  \bibnamefont{et~al.}, \bibinfo{journal}{\pra} \textbf{\bibinfo{volume}{68}},
  \bibinfo{pages}{042319} (\bibinfo{year}{2003}).

\bibitem[{\citenamefont{van Loock et~al.}(2006)\citenamefont{van Loock, Ladd,
  Sanaka, Yamaguchi, Nemoto, Munro, and Yamamoto}}]{van_loock_hybrid_2006}
\bibinfo{author}{\bibfnamefont{P.}~\bibnamefont{van Loock}},
  \bibnamefont{et~al.}, \bibinfo{journal}{Phys. Rev. Lett.}
  \textbf{\bibinfo{volume}{96}}, \bibinfo{pages}{240501}
  (\bibinfo{year}{2006}).

\bibitem[{\citenamefont{Helstrom}(1976)}]{helstrom_quantum_1976}
\bibinfo{author}{\bibfnamefont{C.~W.} \bibnamefont{Helstrom}},
  \emph{\bibinfo{title}{Quantum Detection and Estimation Theory}}
  (\bibinfo{publisher}{Academic Press Inc}, \bibinfo{year}{1976}).

\bibitem[{\citenamefont{Ivanovic}(1987)}]{ivanovic_to_1987}
\bibinfo{author}{\bibfnamefont{I.~D.} \bibnamefont{Ivanovic}},
  \bibinfo{journal}{Physics Letters A} \textbf{\bibinfo{volume}{123}},
  \bibinfo{pages}{257} (\bibinfo{year}{1987}).

\bibitem[{\citenamefont{Dieks}(1988)}]{dieks_overlap_1988}
\bibinfo{author}{\bibfnamefont{D.}~\bibnamefont{Dieks}},
  \bibinfo{journal}{Physics Letters A} \textbf{\bibinfo{volume}{126}},
  \bibinfo{pages}{303} (\bibinfo{year}{1988}).

\bibitem[{\citenamefont{Peres}(1988)}]{peres_to_1988}
\bibinfo{author}{\bibfnamefont{A.}~\bibnamefont{Peres}},
  \bibinfo{journal}{Physics Letters A} \textbf{\bibinfo{volume}{128}},
  \bibinfo{pages}{19} (\bibinfo{year}{1988}).

\bibitem[{\citenamefont{Jaeger and Shimony}(1995)}]{jaeger_optimal_1995}
\bibinfo{author}{\bibfnamefont{G.}~\bibnamefont{Jaeger}} \bibnamefont{et~al.},
  \bibinfo{journal}{Physics Letters A} \textbf{\bibinfo{volume}{197}},
  \bibinfo{pages}{83} (\bibinfo{year}{1995}).

\bibitem[{\citenamefont{Riis and Barnett}(1997)}]{Riis1997}
\bibinfo{author}{\bibfnamefont{E.}~\bibnamefont{Riis}} \bibnamefont{et~al.},
  \bibinfo{journal}{J. Mod. Opt.} \textbf{\bibinfo{volume}{44}},
  \bibinfo{pages}{1061} (\bibinfo{year}{1997}).

\bibitem[{\citenamefont{Huttner et~al.}(1996)\citenamefont{Huttner, Muller,
  Gautier, Zbinden, and Gisin}}]{huttner_unambiguous_1996}
\bibinfo{author}{\bibfnamefont{B.}~\bibnamefont{Huttner}},
  \bibnamefont{et~al.}, \bibinfo{journal}{Phys. Rev. A}
  \textbf{\bibinfo{volume}{54}}, \bibinfo{pages}{3783} (\bibinfo{year}{1996}).

\bibitem[{\citenamefont{Barnett and Croke}(2009)}]{barnett_quantum_2009}
\bibinfo{author}{\bibfnamefont{S.~M.} \bibnamefont{Barnett}}
  \bibnamefont{et~al.}, \bibinfo{journal}{Advances in Optics and Photonics}
  \textbf{\bibinfo{volume}{1}}, \bibinfo{pages}{238} (\bibinfo{year}{2009}).

\bibitem[{\citenamefont{Cook et~al.}(2007)\citenamefont{Cook, Martin, and
  Geremia}}]{cook_optical_2007}
\bibinfo{author}{\bibfnamefont{R.~L.} \bibnamefont{Cook}},
  \bibnamefont{et~al.}, \bibinfo{journal}{Nature}
  \textbf{\bibinfo{volume}{446}}, \bibinfo{pages}{774} (\bibinfo{year}{2007}).

\bibitem[{\citenamefont{Bart\r{u}\v{s}kov\'{a}
  et~al.}(2008)\citenamefont{Bart\r{u}\v{s}kov\'{a}, \v{C}ernoch, Soubusta, and
  Du\v{s}ek}}]{bartkov_programmable_2008}
\bibinfo{author}{\bibfnamefont{L.}~\bibnamefont{Bart\r{u}\v{s}kov\'{a}}},
  \bibnamefont{et~al.}, \bibinfo{journal}{Phys. Rev. A}
  \textbf{\bibinfo{volume}{77}}, \bibinfo{pages}{034306}
  (\bibinfo{year}{2008}).

\bibitem[{\citenamefont{Wittmann et~al.}(2008)\citenamefont{Wittmann, Takeoka,
  Cassemiro, Sasaki, Leuchs, and Andersen}}]{wittmann_demonstration_2008}
\bibinfo{author}{\bibfnamefont{C.}~\bibnamefont{Wittmann}},
  \bibnamefont{et~al.}, \bibinfo{journal}{Phys. Rev. Lett.}
  \textbf{\bibinfo{volume}{101}}, \bibinfo{pages}{210501}
  (\bibinfo{year}{2008}).

\bibitem[{\citenamefont{Chefles and Barnett}(1998)}]{chefles_strategies_1998}
\bibinfo{author}{\bibfnamefont{A.}~\bibnamefont{Chefles}} \bibnamefont{et~al.},
  \bibinfo{journal}{J. Mod. Opt.} \textbf{\bibinfo{volume}{45}},
  \bibinfo{pages}{1295} (\bibinfo{year}{1998}).

\bibitem[{\citenamefont{Fiur\'{a}\v{s}ek and
  Je\v{z}ek}(2003)}]{fiurek_optimal_2003}
\bibinfo{author}{\bibfnamefont{J.}~\bibnamefont{Fiur\'{a}\v{s}ek}}
  \bibnamefont{et~al.}, \bibinfo{journal}{Phys. Rev. A}
  \textbf{\bibinfo{volume}{67}}, \bibinfo{pages}{012321}
  (\bibinfo{year}{2003}).

\bibitem[{\citenamefont{Silberhorn et~al.}(2002)\citenamefont{Silberhorn,
  Ralph, L\"{u}tkenhaus, and Leuchs}}]{silberhorn_continuous_2002}
\bibinfo{author}{\bibfnamefont{C.}~\bibnamefont{Silberhorn}},
  \bibnamefont{et~al.}, \bibinfo{journal}{Phys. Rev. Lett.}
  \textbf{\bibinfo{volume}{89}}, \bibinfo{pages}{167901}
  (\bibinfo{year}{2002}).

\bibitem[{\citenamefont{Lorenz et~al.}(2004)\citenamefont{Lorenz, Korolkova,
  and Leuchs}}]{lorenz_continuous-variable_2004}
\bibinfo{author}{\bibfnamefont{S.}~\bibnamefont{Lorenz}}, \bibnamefont{et~al.},
  \bibinfo{journal}{Applied Physics B} \textbf{\bibinfo{volume}{79}},
  \bibinfo{pages}{273} (\bibinfo{year}{2004}).

\bibitem[{\citenamefont{Lance et~al.}(2005)\citenamefont{Lance, Symul, Sharma,
  Weedbrook, Ralph, and Lam}}]{lance_no-switching_2005}
\bibinfo{author}{\bibfnamefont{A.~M.} \bibnamefont{Lance}},
  \bibnamefont{et~al.}, \bibinfo{journal}{Phys. Rev. Lett.}
  \textbf{\bibinfo{volume}{95}}, \bibinfo{pages}{180503}
  (\bibinfo{year}{2005}).

\bibitem[{\citenamefont{Heersink et~al.}(2006)\citenamefont{Heersink,
  Marquardt, Dong, Filip, Lorenz, Leuchs, and
  Andersen}}]{heersink_distillation_2006}
\bibinfo{author}{\bibfnamefont{J.}~\bibnamefont{Heersink}},
  \bibnamefont{et~al.}, \bibinfo{journal}{Phys. Rev. Lett.}
  \textbf{\bibinfo{volume}{96}}, \bibinfo{pages}{253601}
  (\bibinfo{year}{2006}).

\bibitem[{\citenamefont{Dong et~al.}(2008)\citenamefont{Dong, Lassen, Heersink,
  Marquardt, Filip, Leuchs, and Andersen}}]{dong_experimental_2008}
\bibinfo{author}{\bibfnamefont{R.}~\bibnamefont{Dong}}, \bibnamefont{et~al.},
  \bibinfo{journal}{Nat. Phys.} \textbf{\bibinfo{volume}{4}},
  \bibinfo{pages}{919} (\bibinfo{year}{2008}).

\bibitem[{\citenamefont{Franzen et~al.}(2006)\citenamefont{Franzen, Hage,
  {DiGuglielmo}, Fiurasek, and Schnabel}}]{franzen_experimental_2006}
\bibinfo{author}{\bibfnamefont{A.}~\bibnamefont{Franzen}},
  \bibnamefont{et~al.}, \bibinfo{journal}{Phys. Rev. Lett.}
  \textbf{\bibinfo{volume}{97}}, \bibinfo{pages}{150505}
  (\bibinfo{year}{2006}).

\bibitem[{\citenamefont{Hage et~al.}(2008)\citenamefont{Hage, Samblowski,
  {DiGuglielmo}, Franzen, Fiurasek, and Schnabel}}]{hage_preparation_2008}
\bibinfo{author}{\bibfnamefont{B.}~\bibnamefont{Hage}}, \bibnamefont{et~al.},
  \bibinfo{journal}{Nat Phys} \textbf{\bibinfo{volume}{4}},
  \bibinfo{pages}{915} (\bibinfo{year}{2008}).

\bibitem[{\citenamefont{Marek and Fiurasek}(2009)}]{marek_resources_2009}
\bibinfo{author}{\bibfnamefont{P.}~\bibnamefont{Marek}} \bibnamefont{et~al.},
  \bibinfo{journal}{Phys. Rev. A} \textbf{\bibinfo{volume}{79}},
  \bibinfo{pages}{062321} (\bibinfo{year}{2009}).

\bibitem[{\citenamefont{Lance et~al.}(2006)\citenamefont{Lance, Jeong, Grosse,
  Symul, Ralph, and Lam}}]{lance_quantum-state_2006}
\bibinfo{author}{\bibfnamefont{A.~M.} \bibnamefont{Lance}},
  \bibnamefont{et~al.}, \bibinfo{journal}{Phys. Rev. A}
  \textbf{\bibinfo{volume}{73}}, \bibinfo{pages}{041801}
  (\bibinfo{year}{2006}).

\bibitem[{\citenamefont{Giedke and Cirac}(2002)}]{giedke_characterization_2002}
\bibinfo{author}{\bibfnamefont{G.}~\bibnamefont{Giedke}} \bibnamefont{et~al.},
  \bibinfo{journal}{Phys. Rev. A} \textbf{\bibinfo{volume}{66}},
  \bibinfo{pages}{032316} (\bibinfo{year}{2002}).

\bibitem[{\citenamefont{Takeoka and
  Sasaki}(2008)}]{takeoka_discrimination_2008}
\bibinfo{author}{\bibfnamefont{M.}~\bibnamefont{Takeoka}} \bibnamefont{et~al.},
  \bibinfo{journal}{Phys. Rev. A} \textbf{\bibinfo{volume}{78}},
  \bibinfo{pages}{022320} (\bibinfo{year}{2008}).

\bibitem[{\citenamefont{Wittmann et~al.}(2010)\citenamefont{Wittmann, Andersen,
  and Leuchs}}]{wittmann_preparation_2009}
\bibinfo{author}{\bibfnamefont{C.}~\bibnamefont{Wittmann}},
  \bibnamefont{et~al.}, \bibinfo{journal}{in preparation, (arXiv:1002.0232 [quant-ph])}  (\bibinfo{year}{2009}).

\bibitem[{\citenamefont{Wittmann et~al.}(2009)\citenamefont{Wittmann, Andersen,
  and Leuchs}}]{wittmann_discrimination_2009}
\bibinfo{author}{\bibfnamefont{C.}~\bibnamefont{Wittmann}},
  \bibnamefont{et~al.}, \bibinfo{journal}{J. Mod. Opt. published online
  (arXiv:0905.2496 [quant-ph])}  (\bibinfo{year}{2009}).

\bibitem[{\citenamefont{Heid and L\"{u}tkenhaus}(2006)}]{heid_efficiency_2006}
\bibinfo{author}{\bibfnamefont{M.}~\bibnamefont{Heid}} \bibnamefont{et~al.},
  \bibinfo{journal}{Phys. Rev. A} \textbf{\bibinfo{volume}{73}},
  \bibinfo{pages}{052316} (\bibinfo{year}{2006}).

\bibitem[{\citenamefont{Sych and Leuchs}(2009)}]{sych_coherent_2009}
\bibinfo{author}{\bibfnamefont{D.}~\bibnamefont{Sych}} \bibnamefont{et~al.},
  \bibinfo{journal}{{arXiv:0902.1895} [quant-ph]}  (\bibinfo{year}{2009}).

\bibitem{afek_quantum_2009}%
  \bibinfo {author} {\bibfnamefont{I.}~\bibnamefont{Afek}}, \bibnamefont{et~al.},
  \bibinfo {journal} {Phys. Rev. A} 
  \textbf{\bibinfo {volume} {79}}, \bibinfo {pages} {043830}  (\bibinfo {year} {2009}).

\bibitem[{\citenamefont{Banaszek et~al.}(1999)\citenamefont{Banaszek,
  Radzewicz, W\'{o}dkiewicz, and Krasi\'{n}ski}}]{banaszek_direct_1999}
\bibinfo{author}{\bibfnamefont{K.}~\bibnamefont{Banaszek}},
  \bibnamefont{et~al.}, \bibinfo{journal}{Phys. Rev. A}
  \textbf{\bibinfo{volume}{60}}, \bibinfo{pages}{674} (\bibinfo{year}{1999}).

\end{thebibliography}


\end{document}